\documentclass{article}

\usepackage{arxiv}

\usepackage[utf8]{inputenc} 
\usepackage[T1]{fontenc}    
\usepackage[hidelinks]{hyperref}       
\usepackage{url}            
\usepackage{booktabs}       
\usepackage{amsfonts}       
\usepackage{nicefrac}       
\usepackage{microtype}      
\usepackage{lipsum}		
\usepackage{graphicx}
\usepackage[square,numbers]{natbib}
\usepackage{doi}
\usepackage{authblk}
\usepackage{dblfnote} 

\title{Building a healthier feed: \\Private location trace intersection driven feed recommendations}

\date{September 21, 2023}	

\author[1,2,*]{Tobin South}
\author[3]{Nick Lothian}
\author[1]{Takahiro Yabe}
\author[1,2]{Alex `Sandy' Pentland}
\affil[1]{MIT Connection Science, Massachusetts Institute of Technology, Cambridge, USA}
\affil[2]{MIT Media Lab, Massachusetts Institute of Technology, Cambridge, USA}
\affil[3]{Verida, Adelaide, Australia}
\affil[*]{tsouth@mit.edu}

\renewcommand{\headeright}{}
\renewcommand{\shorttitle}{Private location trace intersection feed}

\hypersetup{
pdftitle={Building a healthier feed: Private location trace intersection driven feed recommendations},
pdfauthor={Tobin South},
}

\begin{document}
\maketitle
\begin{abstract}
The physical environment you navigate strongly determines which communities and people matter most to individuals. These effects drive both personal access to opportunities and the social capital of communities, and can often be observed in the personal mobility traces of individuals. Traditional social media feeds underutilize these mobility-based features, or do so in a privacy exploitative manner. Here we propose a consent-first private information sharing paradigm for driving social feeds from users' personal private data, specifically using mobility traces. This approach designs the feed to explicitly optimize for integrating the user into the local community and for social capital building through leveraging mobility trace overlaps as a proxy for existing or potential real-world social connections, creating proportionality between whom a user sees in their feed, and whom the user is likely to see in person. These claims are validated against existing social-mobility data, and a reference implementation of the proposed algorithm is built for demonstration. In total, this work presents a novel technique for designing feeds that represent real offline social connections through private set intersections requiring no third party, or public data exposure.
\keywords{Social Media Feeds \and Private Data Sharing \and Personal Data Stores 
\and Mobility Data}
\end{abstract}

\noindent \textbf{Note:} This work has been published in the proceedings of SBP-BRiMS 2023 and is posted on arXiv for ease of access. The manuscript can be accessed at Social, Cultural, and Behavioral Modeling. Lecture Notes in Computer Science, vol 14161. Springer, Cham. 16 September 2023. \url{https://doi.org/10.1007/978-3-031-43129-6_6}


\section{Introduction}
Modern social media platforms design feeds to explicitly optimize for user attention, creating negative effects for both users and broader democratic discourse~\cite{Kubin2021TheRO}. These feeds create a feedback loop of preferential attachment to already popular content and users, leading to disproportionally dominant agents~\cite{Lera2020PredictionAP}. To mitigate this rich-get-richer regime, feeds need to focus on promoting low-popularity content as well, but this is not sufficient — you can't just show people low-quality content or users stop engaging. Rather than promoting the content of the largest creators or random content pulled from across the social internet, we want to optimize social capital~\cite{putnam2000bowling}.

Ideally, we want platforms that achieve sustainability and longer-term value by increasing real-world community connections, trust, capacity for collective action (known as bonding social capital), as well the number of trusted links to people in other communities (known as bridging capital). At its core, your feed should be representative of people that matter to you.

The physical spaces we navigate are strong determinants of many aspects of our lives and values, from our wellbeing~\cite{Jaques2015PredictingSH} to our choice of collaborators~\cite{claudel2017exploration} and the opportunities for economic growth we're exposed to~\cite{chetty2015impacts}. Location history is a better predictor of your interests than the demographics used by current matching apps (e.g., dating app, content recommendation, etc.)~\cite{Xiaowen}, and is strongly predictive of which friendships you already have~\cite{Eagle}. Leveraging GPS location data in social media has a long history demonstrating its value, but has been a series of privacy nightmares that require the sharing of this location data to a third party, often to be stored indefinitely and monetized without any visibility or accountability to the user.

We propose a new approach to building social media feeds that are optimized for social capital building via private matching over location histories, where friends are shown in proportion to their potential to build new within-community relationships and to reinforce community social capital, requiring no third-party interactions or capture of data. 

All data sharing is done securely such that no parties ever see anyone else's location data, and only each pair of users can see the size of the location data overlap between them.

This approach is naturally extensible to a huge range of similar matching approaches ranging from shared photo matching, or shared friendship matching (including from existing social networks), to simple interest matching, all without needing trusted third parties or the exposure of any personal information. In doing so, this creates an in-built incentive for users to make personal data accessible (in an encrypted and private manner) to ensure they appear on others' feeds and can participate in the social ecosystem.

In summary, this paper:
\begin{itemize}
\item Presents a novel approach to building a social feed that represents real social relationships via calculations of set intersections between pairs of individuals without the need for a third party or transmission of \emph{any} unencrypted data, ensuring no escape of any private information to anyone.
\item Validates the value of location trace based matching as a measure of friendship on existing data using two different computational examinations, in turn motivating its use in building a feed optimized for social capital building.
\item Prototypes the approach with a reference implementation and demo app to show that this can be done with extremely low computational cost.
\item Proposes and discusses the extensibility of this private set intersection driven feed paradigm to other personal datasets and explores how this creates new social incentives around private data sharing for matching.
\end{itemize}

A longer, more verbose version of this publication with additional figures will be made available online via pre-print servers due to page limits.

\section{A social recommendation algorithm}\label{sec:algorithm}

For two friends Alice \& Bob, the rate at which content is shown on a feed should be proportional to the physical-world relevancy of their ties (here the likelihood they'll be at the same location).

When Bob friends Alice, Bob can make a request to Alice to begin a multistep private set intersection on the mobility traces (location history) of both Bob and Alice. To make matching on this information easier, Alice can provide each point in her GPS trace, $T_A^r$, as a geohash~\cite{niemeyer2008geohash} at an arbitrary resolution, $r$.

Geohash converts GPS coordinates to strings at a varying length of string to provide an arbitration resolution. For example, one can know that a GPS coordinate is inside the MIT Media Lab building using the string `\texttt{drt2yr7x}'. If Alice desires, she can allow others to match on extremely detailed location traces by using a high-resolution geohash in the form of a long string ($r=8$ characters). Or, if Alice decides her mobility trace should not inform her friend's feeds, she can share the geohash at a low resolution, $r=5$, roughly city level. At the extreme, Alice can choose to deny sharing any data at all. However, sharing more information to match with Bob will allow Alice to be seen more often in Bob's feed, creating an incentive to share data at the level of detail one is comfortable with.

Bob can now perform a private set intersection using his original data, choosing the resolution of his mobility data as well. If Alice and Bob plan to share the exact same resolution of location, the task is simple to perform a private set intersection cardinality where Alice acts as a server and Bob acts as a client~\cite{openminedPSI2022}. Alice encrypts each element of her mobility trace $T_A^r$ using a \emph{commutative} encryption scheme, $H(\cdot)^k$, where $k$ is a users secret key, producing a sequence of encrypted geohash strings denoted $\{H(T(i)_A^r)^{k=A}\}_{\forall i}$. These elements are then inserted into a Bloom filter with a chosen false positive rate $e$ and sent to Bob. Similarly, Bob encrypts each element of his sequence with his key to produce  $\{H(T(i)_B^r)^{k=B}\}_{\forall i}$. Bob sends his encrypted sequence back to Alice who can then encrypt it again using her secret key, $\{H(H(T(i)_B^r)^{k=B})^{k=A} \}_{\forall i}$, to send back to Bob. In order to stop Bob from knowing what locations they match on, Alice can shuffle the mobility sequence to ensure that Bob can only see the size of the set intersection, rather than the actual matches. Once returned, Bob uses the commutative properties of the encryption, $\{H(H(T(i)_B^r)^{k=B})^{k=A} \}_{\forall i} = \{H(H(T(i)_B^r)^{k=A})^{k=B} \}_{\forall i}$, to decrypt the returned sequence using his secret key. This now allows Bob to see the size of the overlap between the elements of his sequence,  $\{H(T(i)_B^r)^{k=A}\}$, and Alice's sequence,  $\{H(T(i)_A^r)^{k=A}\}_{\forall i}$, both encrypted using Alice's secret.

This works simply when Alice \& Bob operate at the same resolution $r$. However, there may be instances where either Alice or Bob want to share information at a lower resolution than the other. This adds difficulty, as the encrypted geohashes will not be able to match if $r_A \neq r_B$. To address this, Alice can share her mobility trace at their chosen resolution and below \emph{simultaneously}. Bob should have a choice in how his algorithm matches on different resolutions. If Bob wants to match on only the best possible resolution, then he can perform the cardinal private set intersection repeatedly, starting from his highest resolution down to $r=1$ or until he finds a non-zero overlap. Alternatively, Bob can just perform a set intersection using all his available resolutions in one go, rewarding Alice for sharing more detailed data. To make this feasible, one should limit the resolution to below $r=9$, since GPS's accuracy becomes untenable below the roughly one-meter size of a resolution 9 geohash. 

Adding to this, we can incorporate time into the match. So far we've taken the locations to be a bag-of-words style collection, however matching on time of day is a simple extension by sharing tuples of location-time pairs, where time is binned into hourly segments. 

Computing these private set intersection cardinalities is fast, requiring only three transmissions as serialized protocol buffers, and can be done with existing open source software~\cite{openminedPSI2022}. Once these mobility overlaps are calculated, we can turn to how they can be used to populate a feed.

To echo the physical-world social capital we're hoping to reinforce, we want the time we see someone's content in the feed to be proportional to the time they're likely to see each other in person. At a first pass, it would seem ideal for the probability of Alice being shown in Bob's feed, $P(A)$, to be proportional to the simple overlap of their locations, $P(A) \propto |T_A \cap T_B|$. However, to avoid Alice gaming the system by sharing all, or a large number of possible geohashes, we need to normalize by the size of Alice's location set. Taking a simple proportion, $\frac{|T_A \cap T_B|}{|T_A|}$, would mean that Alice is incentivized to share few locations that are highly likely to match, and penalizes sharing high-quality data. As such, we normalize by $\sqrt{|T_A|}$ to balance these tradeoffs. This gives the simple formula of $P(A) \propto \frac{|T_A \cap T_B|}{\sqrt{|T_A|}}$.

In general, one benefit of this content ranking paradigm is its ability to incentivize people to share data in an attempt to appear on their friend's feeds, helping address the cold-start problem.  However, to facilitate seeing friends who post little or no content, Laplace smoothing could be applied to the ranking weighting.

\subsection{An extendable paradigm}\label{sec:extendable}
This method, although grounded in mobility traces and its value for building social capital, is extremely extensible to other private data. Any common set of interests or activities could be used to build a recommendation system. In particular, where individuals may want to privately hold data and not share it with a wider community is ideal for this paradigm. Examples include: existing friendship ties (which many on social media choose to keep private), which could be used to over-emphasize showing the content of friends who have many mutual friends; a set of private interests (for example, the collection of movies on Netflix one has watched); or a stated set of shared goals (are you looking to find people on this service that are here to make jokes, discuss politics, or look for new friends outside your current circle), without having to publicly declare such a desire. Beyond simple set matching, intersection on more complex data (akin to the problem with GPS locations) can also be addressed.

As an example, take the case of wanting to see posts from friends who have the same (or very similar) photos from parties (often transmitted between friends via AirDrop, messaging platforms, or photo sharing platforms rather than posting publicly) or, in technically identical but somewhat different framing, those with similar memes stored as you. While transmitting these photos to perform set intersection is feasible, minor differences in photos (e.g., cropping, editing, etc) and transmission costs from photo sizes make this suboptimal. Instead, we can use perceptual hashing~\cite{zauner2010implementation} as a stand-in for the geohashing technique. This is an almost exactly analogous approach as the algorithm above, this time just optimizing for shared experiences and memories or shared cultural taste.


This in turn goes a long way to address a fundamental question: do users actually want this feature? 
In a bottom-up social media ecosystem, users have the choice to use, or not use, any given set of features so long as there is a plurality of interoperable interface providers. In general, users may desire not to optimize for social capital, preferring plain entertainment, or purely shared taste as in the above example. While no solution will be possible without any user demand, this approach works to incentivise user uptake of data availability for matching through the desire to appear on friends' feeds, in turn requiring you to share data. 

\subsection{A view towards Web3}\label{sec:web3}
While altering the feeds of traditional social media platforms is challenging by design, the emergence of Web3 social protocols makes building this into a service significantly more feasible. Public social graphs such as the Lens protocol or the AT Protocol by Bluesky Labs allow for direct querying of friends' addresses/handles which could be used to request a private set intersection on mobility data. 
Tying into this, Web3 has presented new paradigms for storing private data~\cite{pentland2021building} that can be requested on-chain through toolkits such as Verida Vault or Disco Data Backpacks, multi-chain protocols for interoperable database storage. In general, any system acting as a personal data store (such as openPDS~\cite{montjoyeOpenPDSProtectingPrivacy2014,de2012trusted}) will enable this approach.

The composable nature of technology in Web3 is oriented towards allowing systems like feeds, content networks, and social graphs to be interoperable and interchangeable. While this is one possible tool in the plural toolkit of social feeds, as we will see in the validation section below, the tool developed here is surprisingly strong for enabling real-world connections and community social capital.

\section{Validation and demonstration}

While, the relationship between mobility traces and friendships are well established~\cite{Eagle,dong2011modeling}, it is important to validate that even after geohashing the locations they remain a predictor of friendship. We draw on three datasets of real human behaviors to demonstrate the performance.

\begin{figure*}[h!]
    \centering
    \includegraphics[width=\textwidth]{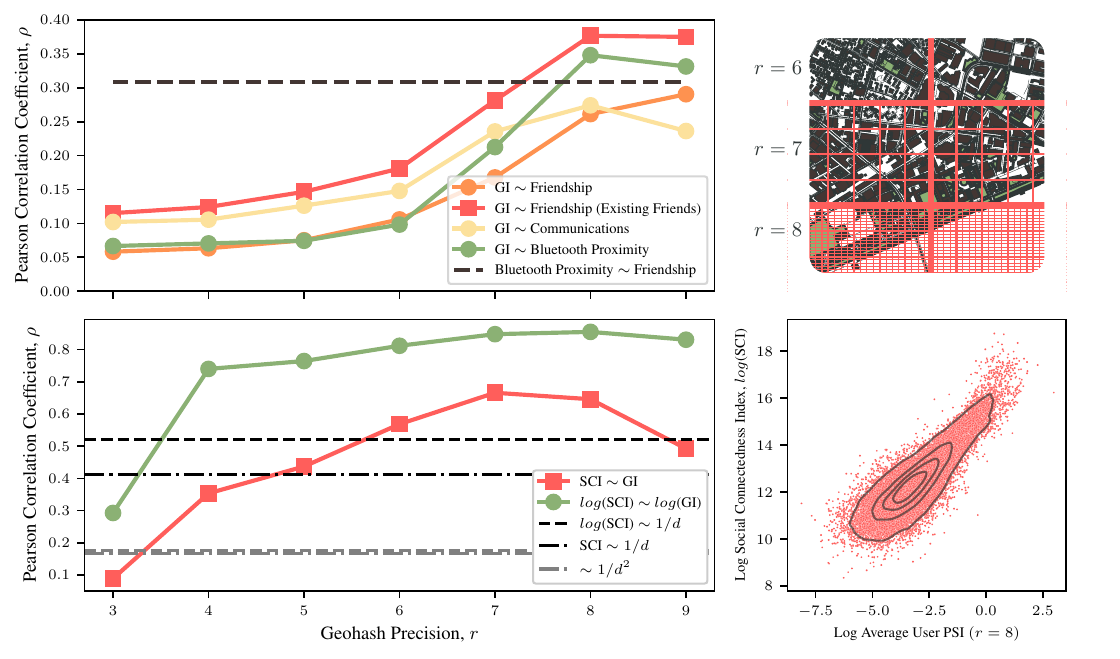}
    \caption{Diverse measures of friendship correlate with set intersections of geohashed location histories between individuals. With sufficient resolution (r), the geohash intersection (GI) between members of a local community~\cite{aharony2011social} is more predictive of self-reported strength of friendship for existing friends and almost as predictive of baseline friendship as bluetooth proximity (top left). More generally, in the bottom row, averages across zip code pairs of all pairwise GI are strongly correlated with Facebook's social connectedness index (SCI)~\cite{SCI} across multiple resolutions, outperforming a zip-zip gravity model.}
    \label{fig:geohash}
\end{figure*}

To show that friendship can be predicted at the individual pair level, we used the Friends and Family experiment where data was collected on 130 residents of a young-family residential living community adjacent to a major research university in North America~\cite{aharony2011social}. This data included GPS coordinates of individuals collected through their mobile phones in addition to Bluetooth proximity between participants' phones in the dataset, call and SMS records between participants, and surveys completed at regular intervals where participants were asked about their perceived strength of friendship with one another (on a scale of zero to seven, where zero is no friendship and 7 is the closest possible friend or partner).

As shown in the top left of \autoref{fig:geohash}, we can geohash these GPS locations at varying levels of resolution to compare how this geohash precision $r$ maps onto reported friendships. To examine this relationship, we apply the same procedure as above, finding set intersection cardinalities between geohashes of individual mobility traces. 
As a baseline, we compare the relationship between Bluetooth proximity, a measure of who you're physically nearby at all times, and the reported friendship. At high resolution, the private set intersection cardinality of your geohashed mobility data is as powerful a predictor of your self-reported friendships as the baseline from Bluetooth proximity; and moreover, if we only examine pairs of individuals who are already friends (reporting any friendship greater than zero), we find that the strength of this friendship can be explained better by your location trace than by your Bluetooth proximity. 

To help explain this, we can examine the relationship between your location intersection with friends and your Bluetooth proximity with them. Even at high resolutions, these are only 35\% correlated. This is largely explained by the time window nature of the algorithm presented in \autoref{sec:algorithm}. The mobility trace is picking up not just seeing one another in person, but sharing a common set of interests as measured by location. We can also see this effect in the relationship between location and communication (measured as the sum of total calls and SMS messages between each pair of individuals) where again, communication does not fully explain friendship.

To ensure this generalizes beyond this small community, we combined two large-scale recent datasets. First, a collection of privacy-preserving mobility traces of over 200K anonymized and opted-in individuals’ mobile phones from five major US metropolitan areas over 2016 and 2017, provided by Cuebiq\footnote{\url{https://www.cuebiq.com/}}. For each pair of zip codes within each metropolitan area, the location geohash intersection is calculated between every combination of individuals who sleep overnight in the zip codes. The average zip-to-zip location intersection is then compared to Meta's publicly available social connectedness index (SCI), a measure of how many Facebook friendships exist between two zip codes~\cite{SCI}.

The results are broadly consistent across the metropolitan areas, with the bottom of \autoref{fig:geohash} showing the relationship between SCI and the geohash intersection (GI) for Boston. The left subfigure shows the correlation of these zip-zip measures across geohash resolutions, with higher resolutions well outperforming a gravity model (both inverse of distance and inverse of distance squared). The rightmost subfigure shows this relationship for a high-resolution geohash with added 2D kernel density mapping. 

\subsection{Reference architecture}
To help demonstrate how this algorithm could be deployed we have implemented a reference architecture built on these principles and created a small demo app to accompany it\footnote{\url{https://github.com/tobinsouth/PSI-Social-Feed}}. 
At a basic level, this app allows a user to enter simple records on the client end and have them matched via a secure private set intersection cardinality with an existing prefilled server. This is done entirely through Javascript, and is an extremely low computational cost operation that can be done quickly on both the client and server side. 
This demo allows for users to match with not only mobility data, but other list-style data as well. 

Much future work exists to build out this algorithm and toolkit into a broader product. In order to effectively interface this algorithm with its use case, you need a feed of content to leverage off of. This could be done by attempting to alter one's existing social feeds (e.g. Twitter) using their APIs or through manipulation of their webpages, but cleaner solutions exist using new Web3 social protocols.

\section{Discussion}
\subsection{Social rankings gamability}
Many existing social and information feeds have resulted in content creators working to `game' the feed through optimizing their content to match the algorithm. 
In contrast, the ranking proposed here deliberately makes it hard to promote oneself more widely than deserved. 

If individuals try to game the system by sharing many location points, they're penalized. Popular influencers can't be everywhere at once to be seen on everyone's feed, and even if a user fabricates a large amount of location data, that would be penalized during normalization. 
Showing few high-resolution places is the best way to appear in the feeds of those you're most likely to bond with. 
As a result, the `fitness' of any given person in the algorithm is fundamentally limited, stopping the possibility of dominant agents emerging~\cite{Lera2020PredictionAP}.

\subsection{Sharing incentives}
This protocol only works as users make their data available for private set matching, creating a cold start problem. Fortunately, as a protocol like this proliferates, users have an incentive to begin sharing their data as soon as their friends do to optimize for exposure on their friends' feeds. 

This cold start problem is not insurmountable. Recent years have seen the rise in alternative social media platforms which have captured mass adoption through creating features that meaningfully consider the negative effects of previously dominant social media offerings (such as BeReal). 

\subsection{Echo-chambers}
By design, this approach to building a social feed reinforces and strengthens existing ties between people and communities. However, this same principle is inherently a driver of echo-chamber formation~\cite{nasim2022we}. In a context where people have shared experiences, context, or location data with people of the same opinions, this algorithm would reinforce ideological segregation and contribute to political polarization. At a fundamental level, this is already at play in the ways we interact with our share physical spaces~\cite{moro2021mobility}; this algorithm may act to reinforce this tendency. 

Ensuring broad exposure to ideas to mitigate political and social fracture is ideal. 
In general, this is far from a solved problem in any social media~\cite{oro72186} but is an important consideration in the design and ultimate rollout of any feed approach.

The proposed Laplace smoothing goes some way to address the echo chamber problem, but we could also draw from existing literature and experience on avoiding polarization and echo-chamber formation. One such approach could be to periodically show content from non-similar users. This could be achieved by effectively adding noise to the recommendation feed; which, while potentially making a less desirable feed for the user, would ensure more diversity of exposure. 

Finally, it is worth noting that the type of homophily being reinforced by this approach is different in character from echo-chamber formation in traditional social media. Traditional social media explicitly ideologically narrows your content based on engagement patterns, allowing some preferences and interests to be exemplified and radicalized as you are presented more concentrated agreeing opinions. In contrast, this approach does not immediately send individuals down specific content `rabbit holes,' potentially leading toward radicalization based only on casual initial interest. In this case, any polarization that occurs happens in step with one's expressed real-world interests and beliefs, and is complementary to physical-world interactions and conversations. 

\section{Conclusion}
This work presents a novel approach to constructing a social feed that represents relationships with friends in terms of real-world quantities beyond social media activity. This is achieved without exposing one's data publicly or relying upon third-party data brokers via private set intersection cardinalities applied to personal datastores. The system supports a variety of data for intersection, focusing firstly on the value of private location traces as tools to measure the likelihood of real-world interaction. Geohashing is used to map real-valued GPS points to strings of arbitrary resolution for matching. The value of location hashes for social capital is validated using existing data on friendships and location, and a demonstration app is built. This work presents an altogether different strategy for populating social feeds by focusing on private data sharing, user consent, and building social capital.

\subsubsection{Acknowledgements} We thank Dinh Tuan Lu for his contributions to the code-base for the demonstration of this idea.

\bibliographystyle{unsrtnat}
\bibliography{references}  
\end{document}